\documentclass[12pt]{article}
\pdfoutput=1

\usepackage[english]{babel}

\usepackage[a4paper,text={16.8cm,22.4cm}]{geometry}
\usepackage{amsmath,amsfonts,slashed,amssymb,tikz,bm,psfrag,graphicx,color,dsfont}
\usepackage{hyperref}
\hypersetup{
    colorlinks=true,
    linkcolor=black,
    citecolor=black,
    filecolor=black,
    urlcolor=black,
}
\RequirePackage[sort&compress,square,comma,numbers]{natbib}
\allowdisplaybreaks 
\usepackage{amsmath}

\begin{document}

\begin{titlepage}

\begin{flushright}
\normalsize
MITP/17-017\\
March 21, 2017
\end{flushright}

\vspace{0.1cm}
\begin{center}
\Large\bf
Exclusive Radiative Decays of Z Bosons \\ in QCD Factorization
\end{center}

\vspace{0.5cm}
\begin{center}
Stefan Alte\footnote{Speaker}$^a$, Yuval Grossman$^b$, Matthias K\"onig$^a$ and Matthias Neubert$^{ab}$\\
\vspace{0.7cm}
{\sl $^a$PRISMA Cluster of Excellence \& Mainz Institute for Theoretical Physics, \\
	Johannes Gutenberg University, 55099 Mainz, Germany\\[3mm]
	$^b$Department of Physics, LEPP, Cornell University, Ithaca, NY 14853, U.S.A.}
 
\end{center}

\vspace{0.8cm}

\abstract{We discuss the very rare, exclusive hadronic decays of a Z boson into a meson and a photon. The QCD factorization approach allows to organize the decay amplitude as an expansion in powers of $\Lambda_{\rm QCD}/m_Z\,$, where the leading terms contain convolutions of perturbatively calculable hard functions with the leading-twist light-cone distribution amplitudes of the meson. We find that power corrections to these leading terms are negligible since they are suppressed by the small ratio $(\Lambda_{\rm QCD}/m_Z)^2\,$. Renormalization-group effects play a crucial role as they render our theoretical predictions less sensitive to the hadronic input parameters which are currently not known very precisely. Thus, measurements of the decays $Z\to M\gamma$ at the LHC or a future lepton collider provide a theoretically very clean way to test the QCD factorization approach. The special case where $M=\eta^({}'{}^)$ is complicated by the fact that the decay amplitude receives an additional contribution where the meson is formed from a two-gluon state. The corresponding branching ratios are very sensitive to the hadronic parameters describing the $\eta-\eta'$ system. Future measurements of these decays could yield interesting information about these parameters and the gluon distribution amplitude. 

}

\vfill

\begin{flushleft}
\small
\sl
38th International Conference on High Energy Physics\\
3-10 August 2016, Chicago, USA
\end{flushleft}

\end{titlepage}

\section{Introduction}
An important challenge to particle physics is the strongly coupled nature of Quantum Chromodynamics (QCD) at low energy scales. For hard exclusive processes with individual hadrons in the final state the QCD factorization approach was developed \cite{Lepage:1979zb,Lepage:1980fj,Efremov:1978rn,Efremov:1979qk,Chernyak:1983ej}. It allows the factorization of short-distance coefficients which can be calculated perturbatively from the hadronic bound-state effects inside the meson which are accounted for by light-cone distribution amplitudes (LCDAs). In this work, we discuss the decays $Z\to M\gamma$ of a Z boson into a meson $M$ and a photon \cite{Grossmann:2015lea,Alte:2015dpo}. The underlying factorization formula for the decay amplitude can be derived employing Soft-Collinear Effective Theory (SCET) \cite{Bauer:2000yr,Bauer:2001yt,Bauer:2002nz,Beneke:2002ph}. This factorization formula is organized as an expansion in terms of $\lambda=\Lambda_{\text{QCD}}/m_Z\,$. Power corrections to the leading terms are suppressed by the small ratio $(\Lambda_{\rm QCD}/m_Z)^2$ and thus negligible, whereas previous applications of the QCD factorization framework suffered from sizeable power corrections (see, e.g. \cite{Beneke:1999br,Beneke:2000ry,Beneke:2001ev,Beneke:2003zv}). Therefore, the decays $Z\to M\gamma$ provide an ideal way to test the QCD factorization approach in a theoretically clean environment. In particular, the case $Z\to \eta^({}'{}^) \gamma$ is interesting due to the complications arising from the flavor-singlet component of the $\eta^({}'{}^)$ meson \cite{Alte:2015dpo}. 

\section{Theoretical framework}
\begin{figure}[h]
\centering
\raisebox{1ex}{\includegraphics[width=0.24\textwidth]{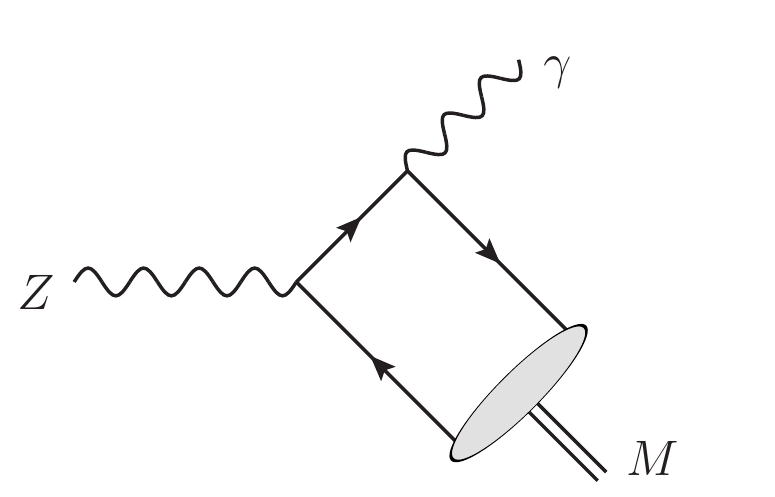}}
\includegraphics[width=0.24\textwidth]{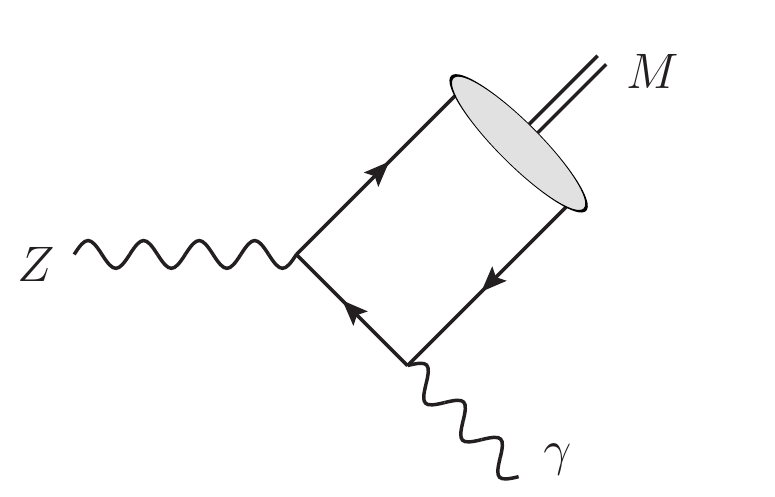}
\includegraphics[width=0.24\textwidth]{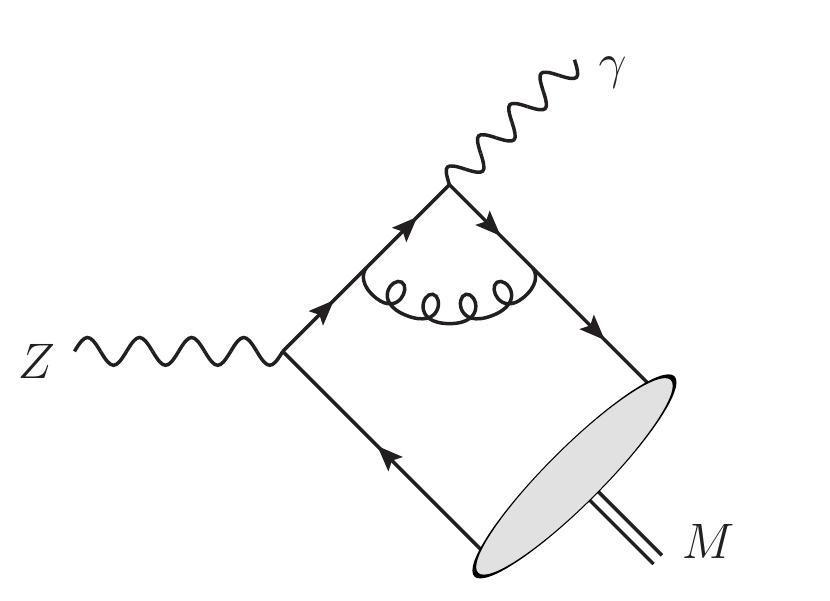}
\includegraphics[width=0.24\textwidth]{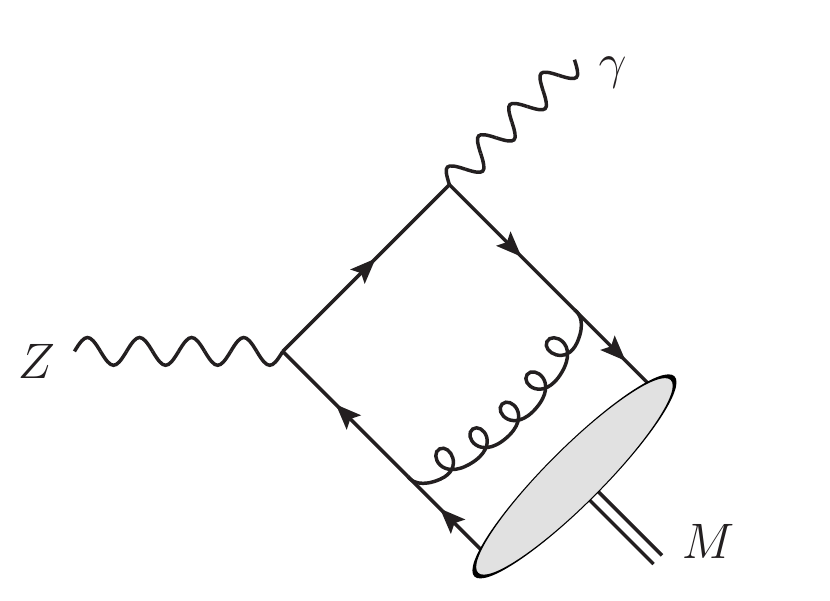}
{\caption{\label{fig:graphs} 
Some Feynman diagrams contributing to $Z\to M\gamma\,$. The meson is shown by the gray blob.}}
\end{figure}

Some representative Feynman diagrams contributing to the decays $Z\to M\gamma$ at leading (LO) and next-to-leading order (NLO) are shown in Figure~\ref{fig:graphs}.
For the case where the final-state meson has a flavor-singlet component in its wavefunction, there exist additional Feynman diagrams where the meson is formed from a two-gluon state. This special case is discussed below. For the derivation of the factorization formula in SCET, we assign the momenta $k$ and $q$ to the final-state meson and photon, respectively. In the rest frame of the decaying Z boson, these momenta fulfill $k^\mu\approx E n^\mu$ and $q^\mu\approx E\bar{n}^\mu\,$, where $E=m_Z/2\,$. $n$ and $\bar{n}$ are two light-like vectors obeying $n\cdot \bar{n}=2\,$. The constituents of the meson can be described by collinear particles in SCET. Their momenta $p_c$ obey the scaling $(n\cdot p_c,\,\bar{n}\cdot p_c,\, p_c^\perp)\sim E(\lambda^2,\,1,\,\lambda)\,$. Exploiting this effective description in SCET, we can derive the factorization formula for the decay amplitude $A$. At the LO in $\Lambda_{\text{QCD}}/m_Z\,$, we find 
\begin{equation}\label{factorization}
   A = -i f_M E \int_0^1\!\text{d}x\,H_M(x,\mu)\,\phi_M(x,\mu) + \mbox{power corrections} \,,
\end{equation}
where $f_M$ is the decay constant of the meson, $H_M$ is the perturbatively calculable hard function, $\mu$ is the factorization scale and $\phi_M$ denotes the leading-twist quark-antiquark LCDA of the meson. 
\par
\begin{figure}[h]
\begin{center}
\includegraphics[height=0.262\textwidth]{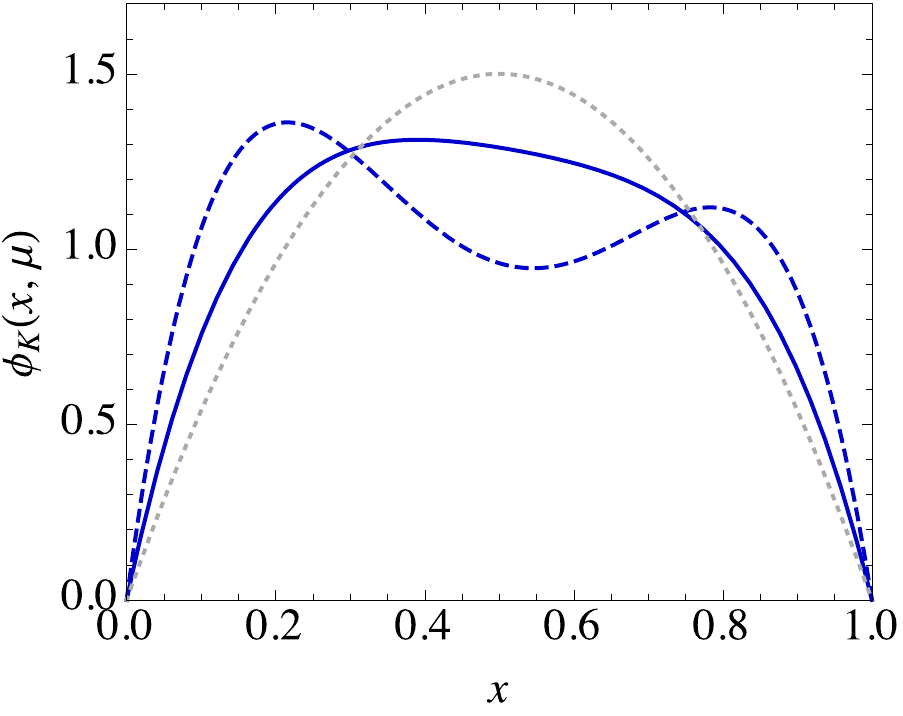}
\includegraphics[height=0.268\textwidth]{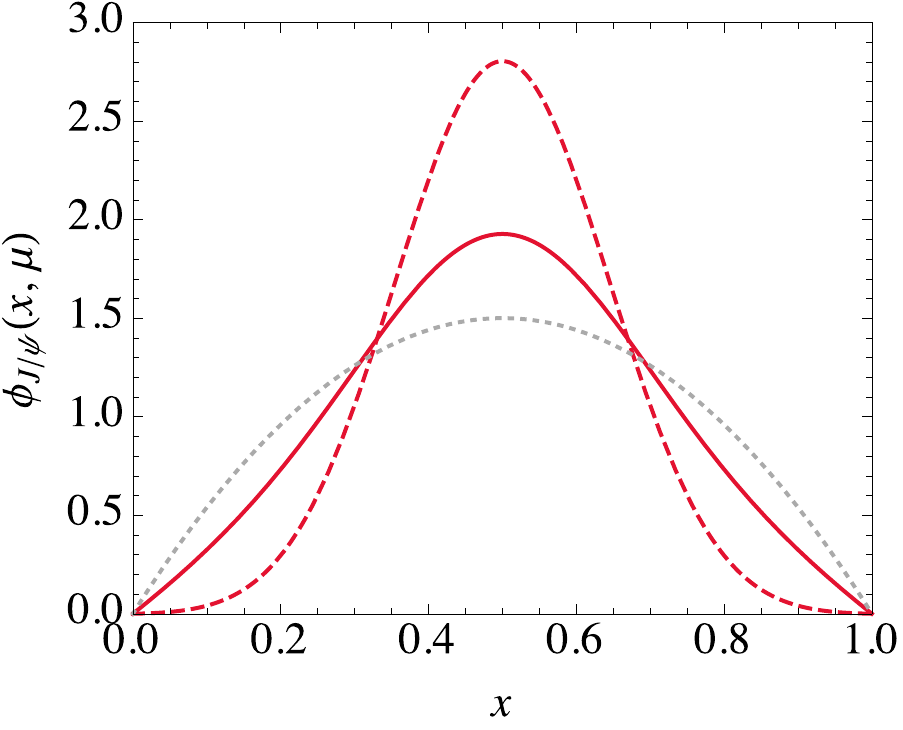}
\includegraphics[height=0.268\textwidth]{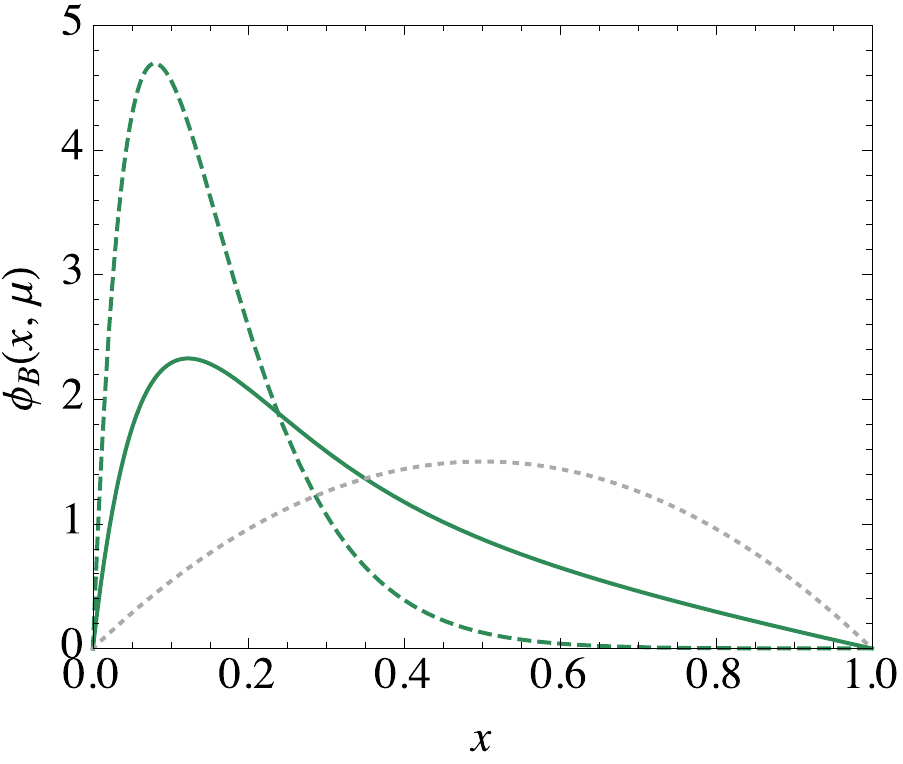}
{\caption{\label{fig:RG} 
Behavior of the LCDAs of the kaon (left), the $J/\psi$ meson (middle) and the $B$ meson (right) under RG evolution from the scale $\mu_0=1$\,GeV (dashed lines) to the scale $\mu=m_Z$ (solid lines). The dotted gray line corresponds to the asymptotic form $6x(1-x)$.}}
\end{center}
\end{figure}

 
As an example, we consider the LCDA for a pseudoscalar meson $P$. It is defined according to
\begin{equation}
\langle P(k)|\,\bar q(t\bar n)\,\frac{\rlap{\hspace{0.02cm}/}{\bar n}}{2}\,\gamma_5\,
    [t\bar n,0]\,q(0) |0\rangle
   = - if_M E \int_0^1\!\text{d}x\,e^{ixt\bar n\cdot k}\,\phi_P(x,\mu) \,,
\end{equation}
where $\bar{q}$ and $q$ are collinear quark spinors, $[t\bar n,0]$ is a Wilson line from $0$ to $t\bar{n}$ and $x$ corresponds to the longitudinal momentum fraction of the quark inside the meson. The leading-twist LCDAs fulfill the expansion \cite{Lepage:1979zb,Chernyak:1983ej}
 \begin{equation}\label{Gegenbauer}
   \phi_M(x,\mu) = 6x(1-x) \left[ 1 + \sum_{n=1}^\infty a_n^M(\mu)\,C_n^{(3/2)}(2x-1) \right] \,,
\end{equation}
where $C_n^{(3/2)}$ are Gegenbauer polynomials with the corresponding Gegenbauer moments $a_n^M\,.$ Since these Gegenbauer moments are non-perturbative input parameters, they have to be to be extracted from experiments or non-perturbative approaches like light-cone QCD sum rules (see e.g. \cite{Ali:1993vd,Ball:1996tb,Ball:1998sk}) or lattice QCD \cite{Arthur:2010xf}. The evolution of the LCDAs from a low hadronic scale $\mu_0=1\text{ GeV}$ up to the factorization scale $\mu= m_Z$ is illustrated for different mesons in Figure~\ref{fig:RG}.
It is important to note that the LCDAs approach their asymptotic form $6x(1-x)$ under renormalization-group (RG) evolution from the low hadronic scale to the factorization scale. Therefore, the high factorization scale makes our predictions less sensitive to the hadronic input parameters.
\par 
Evaluating the contributing Feynman diagrams, we find the decay amplitude
\begin{equation}\label{ampl1}
   i A (Z\to M\gamma)
   = \pm\frac{eg f_M}{2\cos\theta_W} \left[ i\epsilon_{\mu\nu\alpha\beta}\,
    \frac{k^\mu q^\nu\varepsilon_Z^\alpha\,\varepsilon_\gamma^{*\beta}}{k\cdot q}\,F_1^M 
    - \left( \varepsilon_Z\cdot\varepsilon_\gamma^* 
    - \frac{q\cdot\varepsilon_Z\,k\cdot\varepsilon_\gamma^*}{k\cdot q} \right) F_2^M 
    \right] ,
\end{equation}
where the upper (lower) sign corresponds to the case where $M$ is a pseudoscalar (longitudinally polarized) vector meson. It is not possible to produce transversely polarized vector mesons at the LO in the expansion in $\Lambda_{\rm QCD}/m_Z\,$.  The form factors in terms of the Gegenbauer polynomials read
\begin{equation}\label{FVPres}
\begin{aligned}
   F_1^M &= \phantom{-} Q_M \sum_{n=0}^\infty\,C_{2n}^{(+)}(m_Z,\mu)\,a_{2n}^M(\mu) \,, \\
   F_2^M &= - Q_M' \sum_{n=0}^\infty\,C_{2n+1}^{(-)}(m_Z,\mu)\,a_{2n+1}^M(\mu) \,,
\end{aligned}
\end{equation}
where the quantities $Q_M$ and $Q_M'$ involve the couplings of the quarks to the photon and the Z boson. The functions $C_n^{(\pm)}$ are the hard-scattering functions in Gegenbauer moment space. They contain large logarithms of the form $\left[\alpha_s\log\left(m_Z^2/\mu^2\right)\right]^n$ which can be resummed to all orders in perturbation theory using the renormalization group. 
\par
\begin{figure}[h]
\centering
\includegraphics[width=0.33\textwidth]{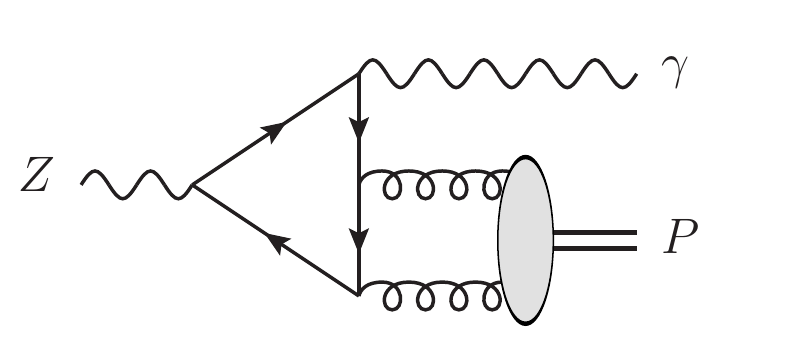}
\includegraphics[width=0.33\textwidth]{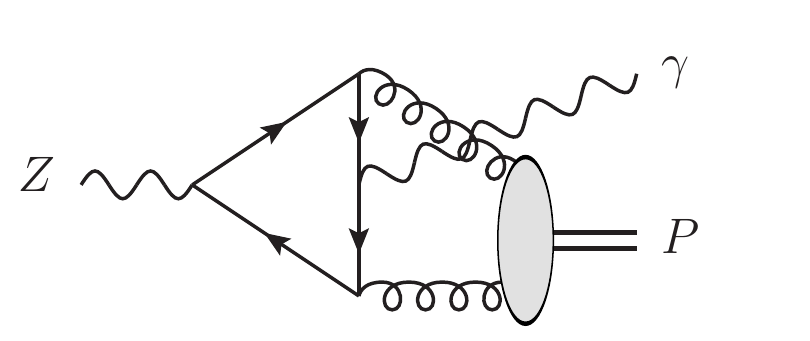}
{\caption{\label{fig:zggg} 
Representative Feynman diagrams which contribute to $Z\to\eta^{(\prime)}\gamma$ at the NLO in perturbation theory. The meson is formed from a two-gluon state.}}
\end{figure}

 
In the special case $Z\to \eta^({}'{}^) \gamma$, the final-state meson contains a flavor-singlet component in its wavefunction. Therefore, the meson can be formed from a two-gluon state at the leading order in $\Lambda_{\rm QCD}/m_Z$ and the NLO in perturbation theory. We show some representative Feynman diagrams in Figure~\ref{fig:zggg}.  
These contributions to the decay amplitude result in additional terms in the factorization formula which contain the leading-twist gluon LCDA of the meson. Furthermore, the RG evolution equations are much more complicated since the quark and the gluon LCDAs mix under RG evolution \cite{Terentev:1980qu,Ohrndorf:1981uz,Shifman:1980dk,Baier:1981pm}. Since the flavor-singlet contains a different number of active quarks at different energy scales between $\mu_0$ and $m_Z\,$, we have to perform a matching at the flavor thresholds in the RG evolution of the hard-scattering coefficients. A final complication is related to the flavor structure of the $\eta^{(\prime)}$ meson. We employ the Feldmann-Kroll-Stech (FKS) scheme \cite{Feldmann:1998vh,Feldmann:1999uf} to describe flavor mixing in the $\eta-\eta'$ system.     

\section{Phenomenological predictions}
\label{sec:flsing}
For the decays where the final-state meson contains no flavor-singlet component in the wavefunction, we present the branching ratios for different decay channels in Table~\ref{tab:BRsZ}.
\begin{table}
\begin{center}
\begin{tabular}{|c|c||c|c|}
\hline 
Decay mode & Branching ratio & asymptotic\ & ~~LO~~ \\ 
\hline 
$Z^0\to\pi^0\gamma$ & $(9.80\,_{-\,0.14}^{+\,0.09}\,{}_\mu\pm 0.03_{f}
 \pm 0.61_{a_2}\pm 0.82_{a_4})\cdot 10^{-12}$ & 7.71 & 14.67 \\ 
$Z^0\to\rho^0\gamma$ & $(4.19\,_{-\,0.06}^{+\,0.04}\,{}_\mu\pm 0.16_{f}
 \pm 0.24_{a_2}\pm 0.37_{a_4})\cdot 10^{-9}$ & 3.63 & 5.68 \\ 
$Z^0\to\omega\gamma$ & $(2.82\,_{-\,0.04}^{+\,0.03}\,{}_\mu\pm 0.15_{f}
 \pm 0.28_{a_2}\pm 0.25_{a_4})\cdot 10^{-8}$ & 2.48 & 3.76 \\ 
$Z^0\to\phi\gamma$ & $(1.04\,_{-\,0.02}^{+\,0.01}\,{}_\mu\pm 0.05_{f}
 \pm 0.07_{a_2}\pm 0.09_{a_4})\cdot 10^{-8}$ & 0.86 & 1.49 \\
$Z^0\to J/\psi\,\gamma$ & $(8.02\,_{-\,0.15}^{+\,0.14}\,{}_\mu\pm 0.20_{f}
 \,_{-\,0.36}^{+\,0.39}\,{}_\sigma)\cdot 10^{-8}$ & 10.48 & 6.55 \\
$Z^0\to\Upsilon(1S)\,\gamma$ & $(5.39\,_{-\,0.10}^{+\,0.10}\,{}_\mu\pm 0.08_{f}
 \,_{-\,0.08}^{+\,0.11}\,{}_\sigma)\cdot 10^{-8}$ & 7.55 & 4.11 \\
$Z^0\to\Upsilon(4S)\,\gamma$ & $(1.22\,_{-\,0.02}^{+\,0.02}\,{}_\mu\pm 0.13_{f}
 \,_{-\,0.02}^{+\,0.02}\,{}_\sigma)\cdot 10^{-8}$ & 1.71 & 0.93 \\
$Z^0\to\Upsilon(nS)\,\gamma$ & $(9.96\,_{-\,0.19}^{+\,0.18}\,{}_\mu\pm 0.09_{f}
 \,_{-\,0.15}^{+\,0.20}\,{}_\sigma)\cdot 10^{-8}$ & 13.96 & 7.59 \\
\hline 
\end{tabular}
{\caption{\label{tab:BRsZ} 
Branching ratios for the decays $Z\to M\gamma\,$. The error budget contains uncertainties from scale variation (subscript ``$\mu$''), the meson decay constant  (``$f$''), the Gegenbauer moments of light mesons (``$a_n$'') and the width parameter in the model for the LCDAs for heavy mesons (``$\sigma$'').
}}
\end{center}
\end{table} 
In the last row $\Upsilon(nS)$ corresponds to a sum over the first three $\Upsilon$ states. The relevant uncertainties arise from the hadronic input parameters, namely the decay constant of the meson, the Gegenbauer moments for light mesons and the width parameter in the model for the LCDA for heavy quarkonia. The branching ratios range from $\text{Br}(Z\to\pi^0\gamma)\sim 10^{-11}$ to $\text{Br}(Z\to J/\psi\gamma)\sim 10^{-7}\,$. We find that calculating the branching ratios using the asymptotic form for the LCDAs yields a good approximation, whereas neglecting NLO QCD corrections results in sizeable deviations.
\par 
For the decays $Z\to\eta^{(\prime)}\gamma$ there exist different sets for the Gegenbauer moments \cite{Agaev:2014wna,Kroll:2013iwa} and the FKS mixing parameters \cite{Feldmann:1998vh,Escribano:2005qq}. Corresponding to these different sets of hadronic input parameters, we find branching ratios which range from $0.1\cdot 10^{-10}$ to $1.7\cdot 10^{-10}$ for the decay $Z\to \eta \gamma$ and from $3.1\cdot 10^{-9}$ to $4.8\cdot 10^{-9}$ for the decays $Z\to \eta'\gamma\,$. The dependence of the branching ratios to the hadronic input parameters is strong such that measurements of the decays $Z\to \eta^({}'{}^) \gamma$ could yield interesting insights to these parameters. In addition, we find that measurements of these decays could be used to directly access the gluon LCDAs.

\section{Conclusions and outlook}
We performed a theoretical analysis of the decays $Z\to M\gamma$ using the QCD factorization approach. These decays provide a theoretically very clean way of testing the QCD factorization approach since power corrections to the leading terms in the factorization formula are negligible. RG evolution plays a crucial role since the high factorization scale gives rise to predictions which are less sensitive to the hadronic input parameters. For the special case $Z\to \eta^({}'{}^) \gamma\,$, we discussed the complications arising from the flavor-singlet component of the $\eta^({}'{}^)$ meson. Measurements of these decays could yield information about the gluon LCDA.  
\par 
In addition to the decays presented here, we studied the decays $W\to M\gamma$ of W bosons in \cite{Grossmann:2015lea} and the decays $h\to M V$ of Higgs bosons, where $V\in\{\gamma,\, Z,\, W\}\,$, in \cite{Koenig:2015pha,Alte:2016yuw}. In particular, the decays $h\to M V$ exhibit highly interesting applications in models beyond the Standard Model (SM). They can be used as powerful probes of some couplings of the Higgs boson to the other SM particles.

\section{Acknowledgments}
S. A. thanks the organizers of this conference for their hospitality. The work of S.A., M.K. and M.N. is supported by the Advanced Grant EFT4LHC of the European Research Council (ERC), the Cluster of Excellence {\em Precision Physics, Fundamental Interactions and Structure of Matter\/} (PRISMA -- EXC 1098), grant 05H12UME of the German Federal Ministry for Education and Research (BMBF), and the DFG Graduate School {\em Symmetry Breaking in Fundamental Interactions\/} (GRK 1581). The work of Y.G. is supported by the U.S. National Science Foundation through grant PHY-0757868 and by the United States-Israel Binational Science Foundation (BSF) under grant no.~2010221.

\end{document}